# Automatic White Blood Cell Measuring Aid for Medical Diagnosis

Pramit Ghosh, Debotosh Bhattacharjee, Mita Nasipuri and Dipak Kumar Basu

*Abstract—* Blood related invasive pathological investigations play a major role in diagnosis of diseases. But in India and other third world countries there are no enough pathological infrastructures for medical diagnosis. Moreover, most of the remote places of those countries have neither pathologists nor physicians. Telemedicine partially solves the lack of physicians. But the pathological investigation infrastructure can't be integrated with the telemedicine technology. The objective of this work is to automate the blood related pathological investigation process. Detection of different white blood cells has been automated in this work. This system can be deployed in the remote area as a supporting aid for telemedicine technology and only high school education is sufficient to operate it. The proposed system achieved 97.33% accuracy for the samples collected to test this system.

## I. Introduction

WHITE blood cells(WBCs) are classified into five major categories, like Lymphocyte , Monocytes, Neutrophils , Eosinophil and basophil [1]. Neutrophils, basophils, and eosinophils have a multi-lobed nucleus. These are differentiated based on the color of the cytoplasm, size and the color of the nucleus. The white blood cell count provides information about various illnesses and also helps to monitor the patient's recovery after initiation of treatment. One measure, the differential blood count, indicates the type of blood cells which are most affected. The normal white blood cell count is between 4500 and 10000 cells per micro-litre depending on the sex and age of the individual with a composition[2] of Neutrophils: 50 – 70%; Lymphocytes: 25 – 35%; Basophils: 0.4 – 1%; Eosinophils: 1 – 3%; Monocytes: 4 – 6%; as shown in figure 1.

A high blood count (above 30000 cells per micro-litre) does not indicate any specific disease but indicates infection, systemic illness, inflammation, allergy, leukemia and tissue injury caused due to burns [3]. The count of white blood cells also increases when certain medicines like antibiotics or anti-seizure drugs are applied. Smoking and too much of mental stress also increases the count of the white blood cells in the body. Moreover, when the count of white blood cell is on the higher side, the risk of cardiovascular mortality also increases. It turns into a vicious cycle. On the other hand, a low count of white blood cells indicates viral infections, low immunity and bone marrow failure [3]. A severely low white blood count that is the count of less than 2500 cells per micro-litre is a cause for a critical alert and possesses a high risk of sepsis [4].

In conventional procedure, glass slides containing blood samples are dipped into Lisman solution before placing it into microscope [5]. Microscope enlarges the pictures of blood samples for manual detection of different white blood cells; but this manual process totally depends on pathologist. Some auto cell counting units exist like Cellometer [15], automatic blood cell counter (BL500)[16] etc. But costs of those machines are much higher. As a matter of fact, it is difficult to place such auto WBC counting machine in every health care unit. The objective of this work is to design a low cost device which can count WBCs in a given blood sample accurately and efficiently.

The rest of the paper is arranged as follows: section II describes the proposed system; section III presents the results and performance of the system and section IV concludes the work along with discussions on the future scope of this work.

## II. System Details

The entire work is an amalgamation of mechanical engineering, control engineering, image processing, pattern recognition, and fuzzy logic. The system is explained with the help of a block diagram, shown in figure 2, and all the steps of the system are described next.

Pramit Ghosh is with the Department of Computer Science. & Engineering., RCC Institute of Information Technology, Kolkata 700015, India (Email pramitghosh2002@yahoo.co.in).

Debotosh Bhattacharjee, Mita Nasipuri and Dipak Kumar Basu are with the Department of Computer Science. & Engineering. Jadavpur University, Kolkata 700032 India. (Ph +913325735112 Email debotoshb@hotmail.com, mitanasipuri@gmail.com, dipakkbasu@gmail.com).

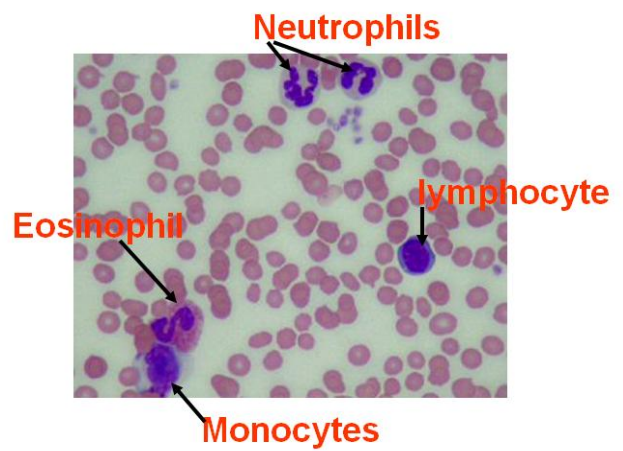

Fig. 1. Different types of whit blood cells.

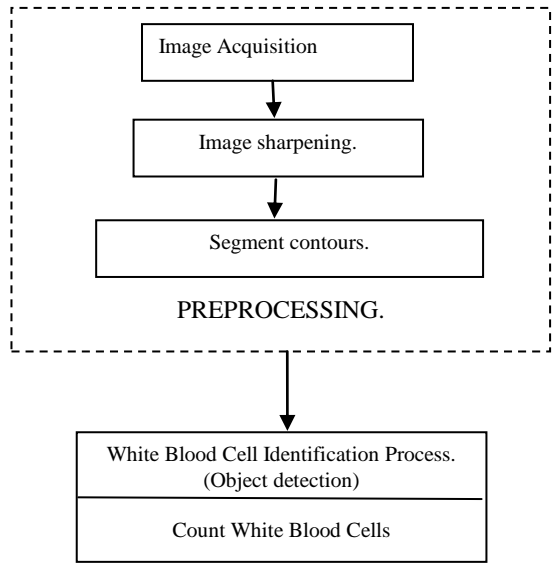

Fig. 2. Block diagram of the system.

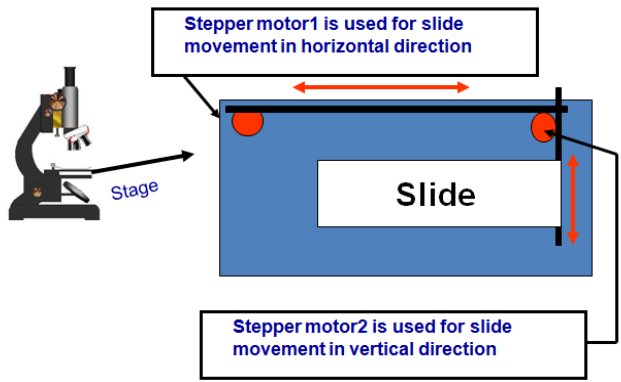

Fig. 3. Slide control through stepper motor.

*A. Image Acquisition*

Image Acquisition is the first step of the system. Digitized images of the blood samples on the slides are acquired with a CCD [6] camera which is mounted upon the microscope. For getting multiple images of a single sample, the glass slide movement is required and it is controlled by two stepper motors [7] in the horizontal and vertical direction shown in Fig 3.

*B. Image Enhancement*

The images obtained from the CCD camera are not of good quality. Laplacian Filter [8] is used to sharpen the edges of the objects in the image.

The Laplacian value for a pixel is denoted by $\nabla^2 f(x,y)$ and it is defined as.

$$\nabla^2 f(x,y) = \frac{\partial^2}{\partial x^2} f(x,y) + \frac{\partial^2}{\partial y^2} f(x,y) \quad (1)$$

This expression is implemented at all points (x,y) of the image through convolution. The Laplacian filter is applied separately on Red, Green and Blue components of the colour images obtained from the CCD camera.

*C. Segmentation*

It is known that only white blood cell has nucleus and red blood cell has no nucleus. In microscopic images the nucleus is stained with blue color, as shown in Fig 1.

The next step is to extract colour information. But the problem is that RGB component contains not only the colour information but also the colour intensity i.e. the RGB values are different for light blue, dark blue and navy blue. So from RGB components it is very difficult to identify the colour. To overcome this problem HSI [9] colour format is used, where H stands for Hue (Pure colour), S is for saturation, i.e. the degree of dilution of pure colour white light and I is for intensity (Gray level)

The RGB to HSI colour space conversion process [10] is performed using the equations (2) – (4).

$$H = \begin{cases} \theta & \text{if } B \leq G \\ 360 - \theta & \text{otherwise} \end{cases} \quad (2)$$

$$\theta = \cos^{-1}\left\{ \frac{\frac{1}{2}[(R-G)+(R-B)]}{\sqrt{[(R-G)(R-G)+(R-B)(G-B)]}} \right\}$$

$$S = 1 - \frac{3}{(R+G+B)}[\min(R,G,B)] \quad (3)$$

$$I = \frac{1}{3}(R+G+B) \quad (4)$$

By analyzing statistical data it is inferred that reddish portion of the image has very low "Hue" value whereas blue has more "Hue" value. To detect the white blood cells, which is stained with blue colour, a high pass image filter is applied on the "Hue" image component of the input color image because "Hue" value of blue is greater than "Hue" value of red.

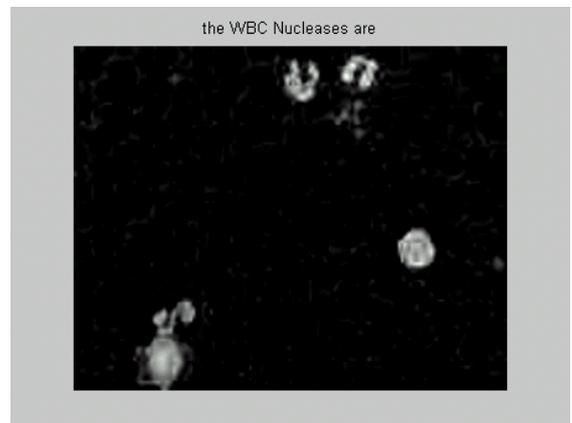

Fig. 4. The stained nucleuses are selected from "Hue" component.

The output of the high pass filtered Hue image is shown in Fig. 4. After that, to obtain the positions of different white blood cell contours, the high pass filtered image is converted into binary image[11] using a threshold value shown in figure. 5. The threshold value calculation [12] algorithm is given next.

Algorithm 1: Calculation of threshold value.

This algorithm is used to calculate the threshold value which will be used to convert the gray image into binary image.

Step 1: Select an initial estimate for T (T=threshold value). The initial value for T is the average gray level of the image
Step 2: Segment the images using T. This will produce two groups of pixels: consisting of all the pixels with gray level values > T called as G1 and consisting of pixels with values <= T called as G2.
Step 3: Compute the average gray level values µ1 and µ2 for the pixels in regions G1 and G2.
Step 4: Compute a new threshold value:
T = 0.5 * (µ1+ µ2).
Step 5: Repeat steps 2 through 4 until the difference in T in successive iterations is smaller than a predefined parameter $T_0$.
Step 6: Stop.

In the binary image some unnecessary contours exist; which are actually noise. After the individual contours in a binary image are segmented by label matrix technique they are checked whether they represent valid contours or noise. This is done by the following algorithm.

Algorithm 2: To find out a valid contours.

This algorithm is used to find out valid contours of the binary image shown in fig 5.

Step1: Apply label matrix technique in the binary image, and store the output matrix in LB variable. LB will have the same dimension as the binary image. The value '1' in each contour in the binary image will be replaced by an integer number in the LB.
Step 2: count = maximum integer value stored in LB; So "count" will contain the number of contours in the input image.
Step 3: index = 1
Step 4: val = maximum gray level value of the pixels in the high pass filtered image.
Step 5 : Find the coordinates of the pixels of the LB where value of the pixel == index;
Step 6: local_Value = maximum gray level value of the pixels in the high pass filtered image whose coordinates are selected in step 5.
Step 7: if local_Value > 0.85* val
  the contour is valid and segment the contour for processing.
  Else
  Not a valid contour
Step 8 : index = index + 1;
Step 9: Repeat step 5 to step 8 until index > count
Step 10 : Stop

The outcome of the algorithm 2 is shown in fig. 6 which is a valid contour. In some situation two white blood cells are overlapped upon each other, on that case two nucleuses are in same contour so it is very important to ensure that one contour must have only one nucleus. The algorithm 3, given next, is used to do this.

Algorithm 3: To separate the overlapped nucleuses.

Two blood cells are in touch with each other, forming a large segment, now to separate those nucleuses the algorithm is given.

Step 1: Dilating the segmented contour, obtained by applying algorithm 2, and subtract the segmented contour pixel-wise from dilated contour. Fig 7 shows the result. This subtracted binary image is actually the mask to get the pixels just outside the nucleus of white blood cell.
Step 2: Find out pixels from the image obtained after high pass filtering lies outside nucleus using the mask obtained from the step 1.
Step 3: value = average value of the pixel values obtained in the step 2.
Step 4: Build an image with the same dimension of the original image; where all the pixels are filled with "value" except the coordinates of the mask; these points are filled with the value of the high pass filtered image shown in Fig 8.
Step 5: Convert the newly created image into binary image using threshold value. This threshold is calculated by the algorithm 1 described earlier. Fig 9 shows the actual non overlapped contour.
Step 6:Segment the contour using label matrix technique.
Step 7: Now each contour contains only one nucleus.
Step 8: Send all contours into object recognition module.
Step 9: Stop.

D. *Object Detection*

The final step is object detection. As discussed earlier, in introduction section, white blood cells are classified into five major categories based on the nucleus shape and colour of the cytoplasm. The nucleus of Monocytes is oval; Lymphocytes is circular and multi-lobe nucleus for Neutrophils, Basophils, and Eosinophils

The height and width of the Monocyte segment has huge difference, whereas in case of Lymphocytes the difference is less. Using this information Monocytes and Lympocytes are identified. These are shown in Fig 10 and Fig 11 separately.

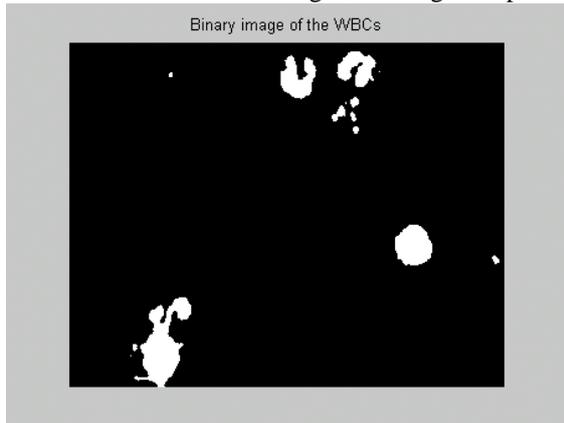

Fig. 5. Binary image of the high pass filtered image (fig 4).

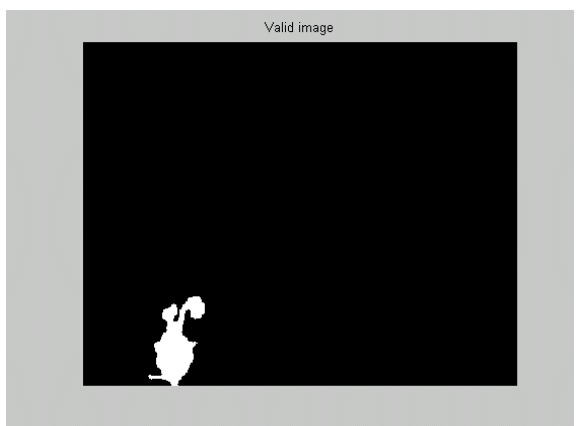

Fig. 6. Selected valid contour.

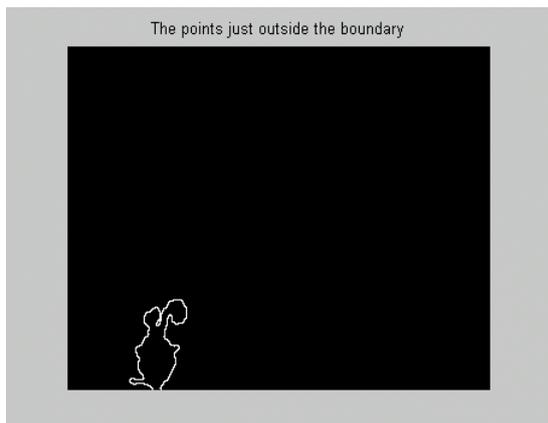

Fig. 7. The mask to find out the points just outside the contour.

In the multi-lobe nucleus segment, the center position is outside the contour. The center point can be easily calculated by applying average calculation technique applied upon the pixels containing nucleus. Based on this technique, the Neutrophils, Basophils, Eosinophils are separated from Monocytes and Lympocytes. Moreover Neutrophils, Basophils and Eosinophils are distinguished based on the cytoplasm colour. Eosinophils have red cytoplasm, for Neutrophils it is white and blue for Basophils.

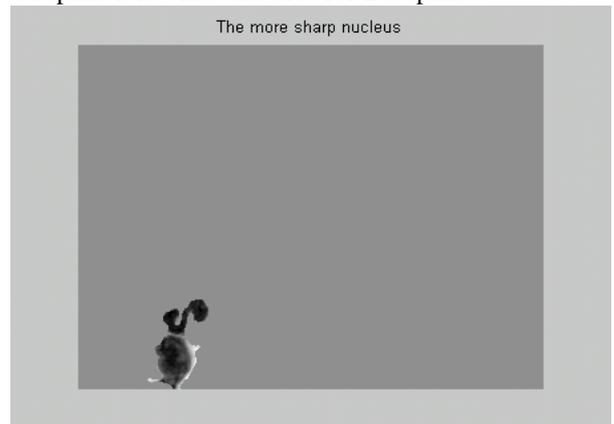

Fig. 8. The image obtain after step 4 in algorithm 3.

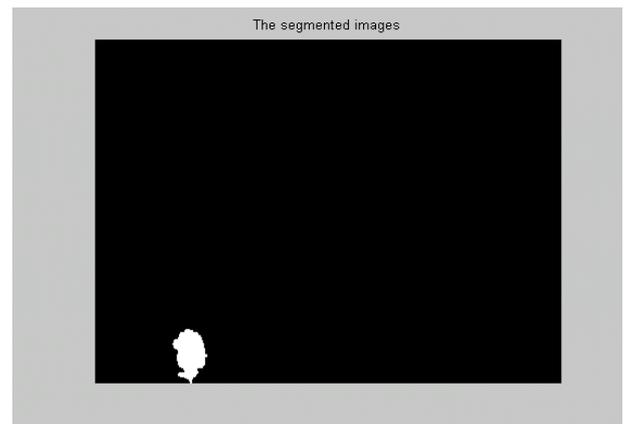

Fig. 9. The segmented nucleus.

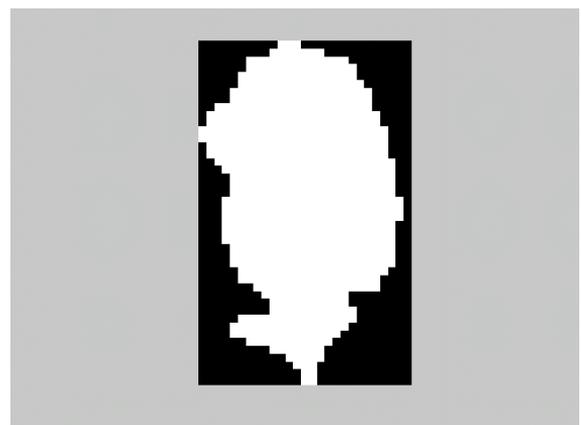

Fig. 10. The Monocytes.

The next section will discuss an algorithm designed to find out white blood cells based on the cytoplasm colour.

Algorithm 4: To distinguish Neutrophils, Basophils and Eosinophils from each other.

This algorithm tests the colour of the cytoplasm to make decisions.

Step 1: The binary image segment contains the mask of the nucleus. Dilate [12] the mask and store the new mask in "Thick_Mask".

Step 2: Subtract segmented mask from "Thick_Mask" and store newly created mask in "Cytoplasm_Mask" shown in Fig 12

Step 3 Get the colour information of the desired cytoplasm area by pixel wise multiplying the enhanced colour image discussed in section B with the "Cytoplasm_Mask"

Step 4: Convert the RGB colour information, of the cytoplasm obtained from step 3, into Hue, and saturation using equation 2 and 3.

Step 5: Assign average Hue and saturation values, generated by the step 4, into Ave_Hue and Ave_Saturation respectively.

Step 6 if Ave_Saturation is very high then
    Cell is Neutrophils because cytoplasm is white.
    Else
    Determine the colour of the cytoplasm by applying fuzzy logic [13] based colour detection system, where fuzzy membership functions are created using equation 5 and colour training sets [10]. After that if the decided color is red then the cell is Eosinophils and if it is blue then Basophils

Step 7 Stop.

$$f(X) = \begin{cases} 0 & \text{if } X < \min(DataSet) \text{ or} \\ & X > \max(DataSet) \\ \sum_{i=1}^{n} Y_i \cdot \prod_{j=1, j \neq i}^{n} \frac{(X - X_i)}{(X_i - X_j)} & ....otherwise \end{cases} \quad (5)$$

where f(X) is the response of the membership function, X is the input value, DataSet is the training set, n is the number of data in the DataSet, each element of the dataset is a pair of values ($X_i$, $Y_i$) where $X_i$ stands for input and $Y_i$ for corresponding output of the membership function.

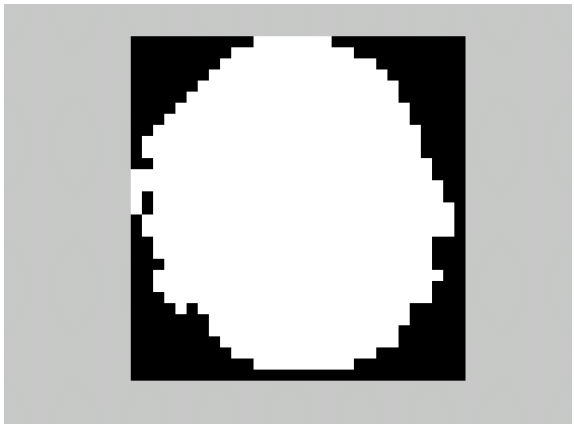

Fig. 11. The Lymphocyte.

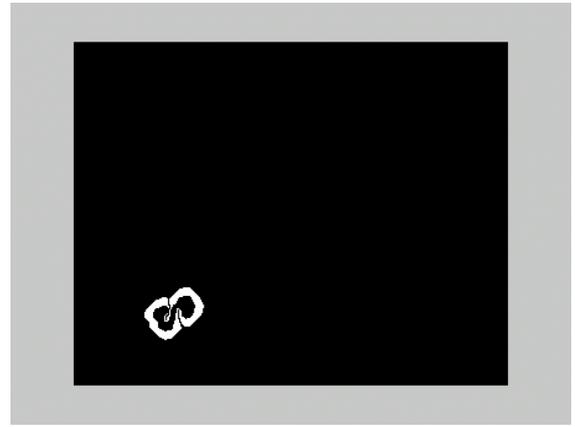

Fig. 12. The cytoplasm detection mask .

III. SIMULATION AND RESULTS

For simulation MatLab 7.1 [14] is used. The image shown in Figure 1 is fed as an input. After analysis, system finds that there are five white blood cells, out of those two are Neutrophils, one is Eosinophils, one is Monocytes and last one is Lymphocytes. The snapshot of the output is shown in figure 13.

For another test image the result is also satisfactory. The input image is shown in figure 14 and corresponding output is shown in figure 15 where number of white blood cells is two, out of which one is Neutrophils, and other is Monocytes.

Fig. 13 The simulation result of the proposed system.

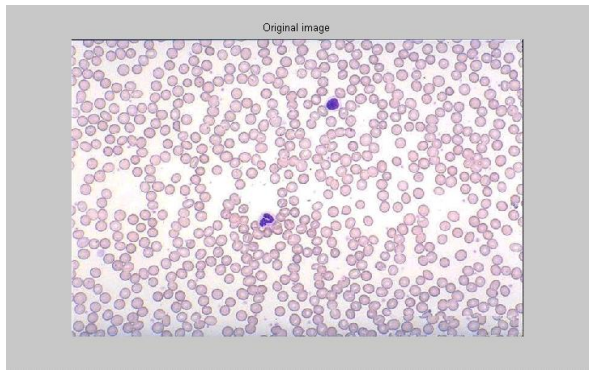

Fig. 14. The second input image set.

Fig. 15. The simulation result for second input image set.

For testing purpose 150 samples are used in which 146 give accurate results. So the accuracy of the system is 97.33%

IV. CONCLUSION

In this paper one novel approach for detection and counting of different types of white blood cells is proposed. This system is cheaper than other auto recognizer like Cellometer, automatic blood cell counter (BL500) etc. This system is easy to install and use. Therefore, it can be installed by people of remote places with very basic level of education. It can reduce the probability of wrong diagnosis in comparison to manual counting. As a next phase authors are trying to automate other microscope-based pathological investigations like detection of bacteria or protozoa.

ACKNOWLEDGMENT

Authors are thankful to Dr. Abhijit Sen for providing pathological data and the "DST-GOI funded PURSE programme", at Computer Science & Engineering Department, Jadavpur University, for providing infrastructural facilities during progress of the work.